\documentclass{conm-p-l}

\usepackage{graphics}

\newtheorem{theorem}{Theorem}[section]
\newtheorem{lemma}[theorem]{Lemma}

\theoremstyle{definition}
\newtheorem{definition}[theorem]{Definition}

\theoremstyle{remark}

\numberwithin{equation}{section}

\newcommand{\F}{{\mathcal F}}
\newcommand{\e}{\epsilon}

\newcommand{\vphi}{\varphi}

\newcommand{\A}{{\mathcal A}}

\renewcommand{\k}{\kappa}
\newcommand{\ga}{\gamma}
\newcommand{\Ga}{\Gamma}

\newcommand{\tS}{\tilde{S}}

\newcommand{\dl}{\delta}

\renewcommand{\th}{\theta}
\newcommand{\Th}{\Theta}
\newcommand{\ra}{\rightarrow}

\newcommand{\Sg}{\Sigma}

\newcommand{\pa}{\partial}

\newcommand{\La}{\Lambda}

\newcommand{\la}{\lambda}

\newcommand{\Om}{\Omega}

\newcommand{\tE}{\tilde{E}}
\newcommand{\tv}{\tilde{v}}
\newcommand{\tth}{\tilde{\theta}}

\def\maprightu#1{\smash{
    \mathop{\longrightarrow}\limits^{#1}}}
\def\maprightd#1{\smash{
    \mathop{\longrightarrow}\limits_{#1}}}
\def\mapdownl#1{
    \llap{$\vcenter{\hbox{$\scriptstyle#1$}}$}\Big\downarrow}
\def\mapdownr#1{\Big\downarrow
    \rlap{$\vcenter{\hbox{$\scriptstyle#1$}}$}}

\begin{document}

\title{Chaos and Shadowing Around a Homoclinic Tube}

\author{Yanguang (Charles)  Li}
\address{Department of Mathematics, University of Missouri, 
Columbia, MO 65211}
\curraddr{}
\email{cli@math.missouri.edu}
\thanks{}

%\author{Author Two}
%\address{Mathematical Research Section, School of Mathematical Sciences,
%Australian National University, Canberra ACT 2601, Australia}

%    General info
\subjclass{37, 35, 34}
\date{}



\begin{abstract}
Let $F$ be a $C^3$ diffeomorphism on a Banach space $B$. $F$ has 
a homoclinic tube asymptotic to an invariant manifold. Around the 
homoclinic tube, Bernoulli shift dynamics of submanifolds is 
established through shadowing lemma. This work removes an 
uncheckable condition of Silnikov [Equation (11), page 625 of 
L. P. Silnikov, Soviet Math. Dokl., vol.9, no.3, (1968), 624-628]. 
Also, the result of Silnikov does not imply Bernoulli shift 
dynamics of a single map, rather only provides a 
labeling of all invariant tubes around the homoclinic tube. The 
work of Silnikov was done in ${\mathbb R}^n$, and the current 
work is done in a Banach space.
\end{abstract}

\maketitle

%\section*{}
%This is an example of an unnumbered first-level heading.

%\specialsection*{This is a Special Section Head}
%This is an example of a special section head%
%%%%%%%%%%%%%%%%%%%%%%%%%%%%%%%%%%%%%%%%%%%%%%%%%%%%%%%%%%%%%%%%%%%%%%%%
%\footnote{Here is an example of a footnote. Notice that this footnote
%text is running on so that it can stand as an example of how a footnote
%with separate paragraphs should be written.
%\par
%And here is the beginning of the second paragraph.}%
%%%%%%%%%%%%%%%%%%%%%%%%%%%%%%%%%%%%%%%%%%%%%%%%%%%%%%%%%%%%%%%%%%%%%%%%

%\section{This is a numbered first-level section head}
%This is an example of a numbered first-level heading.

%\subsection{This is a numbered second-level section head}
%This is an example of a numbered second-level heading.

%\subsection*{This is an unnumbered second-level section head}
%This is an example of an unnumbered second-level heading.

%\subsubsection{This is a numbered third-level section head}
%This is an example of a numbered third-level heading.

%\subsubsection*{This is an unnumbered third-level section head}
%This is an example of an unnumbered third-level heading.

\section{Introduction}

In \cite{Sil68}, Silnikov introduced the concept of a homoclinic tube 
which can be obtained through a transversal intersection of the 
center-unstable and center-stable manifolds of a normally hyperbolic 
invariant manifold under a map in ${\mathbb R}^n$. Intuitively 
speaking, a homoclinic tube can be regarded as a homoclinic orbit on 
which points are replaced by submanifolds. Under certain assumption 
[Equation (11) on page 625 of \cite{Sil68}] which is uncheckable, all 
the invariant tubes in the neighborhood of the homoclinic tube can be 
labeled symbolically. Such a symbolic labeling does not imply Bernoulli 
shift of a single map. The result was proved through a contraction map 
argument on a sequence of metric spaces. In the current article, we 
will adopt a different approach developed in \cite{Li03}. We will 
establish Bernoulli shift dynamics of submanifolds through shadowing 
lemma. The uncheckable assumption of Silnikov is removed. We will work 
in a Banach space, while Silnikov worked in ${\mathbb R}^n$.

Especially for high dimensional systems, homoclinic tubes are more 
dominant structures than homoclinic orbits. In fact, the invariant manifold 
that the homoclinic tube is asymptotic to, can contain smaller scale 
chaotic dynamics as discussed in \cite{Li99} \cite{Li03a}. Although 
structures in a neighborhood of a homoclinic orbit have been extensively 
and intensively investigated, structures around a homoclinic tube have 
not been well studied \cite{Sil68} \cite{Li99} \cite{Li03a}. I believe 
that homoclinic tubes will play an important role in the theory of 
chaos in Hamiltonian partial differential equations.

The article is organized as follows: section 1 is the introduction, 
section 2 contains the setup and definitions, section 3 deals with 
Fenichel fiber coordinates and a $\la$-lemma, and section 4 deals with 
shadowing lemma and chaos.

\section{The Setup and Definitions}

The setup is as follows:
\begin{itemize}
\item {\bf (A1)}. Let $B$ be a Banach space on which a $C^3$-diffeomorphism 
$F$ is defined. There is a normally (transversally) hyperbolic invariant 
$C^3$ submanifold $S$. Let $W^{cu}$ and $W^{cs}$ be the $C^3$ 
center-unstable and center-stable manifolds of $S$. There exist a $C^2$ 
invariant family of $C^3$ unstable Fenichel fibers 
$\{ \F^u(q)\ : \ q \in S \}$ and a $C^2$ invariant family of $C^3$ 
stable Fenichel fibers $\{ \F^s(q)\ : \ q \in S \}$ inside $W^{cu}$ 
and $W^{cs}$ respectively, such that
\[
W^{cu}=\bigcup_{q \in S} \F^u(q)\ , W^{cs}=\bigcup_{q \in S} \F^s(q)\ .
\]
There are positive constants $\k$ and $C$ such that
\begin{eqnarray}
& & \| F^{-n}(q^-)-F^{-n}(q)\| \leq C e^{-\k n} \| q^--q\|\ , \ \ 
\forall n \in {\mathbb Z}^+\ ,  \ \ \forall q \in S \ ,  \ \ \forall q^- 
\in \F^u(q)\ , \nonumber \\
& & \| F^{n}(q^+)-F^{n}(q)\| \leq C e^{-\k n} \| q^+-q\|\ , \ \ 
\forall n \in {\mathbb Z}^+\ ,  \ \ \forall q \in S \ ,  \ \ \forall q^+ 
\in \F^s(q)\ , \nonumber \\
& & \| F^{n}(q_1)-F^{n}(q_2)\| \leq C e^{\k_1 |n|} \| q_1-q_2\|\ , \ \ 
\forall n \in {\mathbb Z}\ ,  \ \ \forall q_1, q_2 \in S \ ,  \nonumber
\end{eqnarray}
where $\k_1 \ll \k$, for example, $\k_1 < \k /300$. $W^{cu}$ and $W^{cs}$ 
intersect along an isolated transversal homoclinic tube $\xi$ asymptotic 
to $S$. $\xi = (\cdots S_{-1} S_0 S_{1} \cdots )$ where $S_j = F^j S_0$, 
$\forall j \in {\mathbb Z}$, and $S_0$ is $C^3$. $\forall j \in {\mathbb Z}$
and $\forall q_j \in S_j$, $q_j$ is on a unique stable fiber $\F^s(q_+)$, 
$q_+ \in S$, and a unique unstable fiber $\F^u(q_-)$, $q_- \in S$. We 
denote such correspondences by $\vphi^+_j$ and $\vphi^-_j$ respectively, 
$\vphi^\pm_j(S) = S_j$. $\vphi^\pm_j$ are $C^2$ diffeomorphisms. Let
\[
\| \vphi^\pm_j - \ \mbox{id} \ \|_{C^1}=\sup_{q \in S} \{ \max \{ \| 
\vphi^\pm_j (q)-q\|, \ \| D \vphi^\pm_j (q) - \ \mbox{id} \ \| \} \} 
\]
where $\mbox{id}$ is the identity map and $D \vphi^\pm_j$ denotes the 
differential of $\vphi^\pm_j$. As $j \ra +\infty$,
\[
\| \vphi^+_j - \ \mbox{id} \ \|_{C^1} \ra 0\ .
\]
As $j \ra -\infty$,
\[
\| \vphi^-_j - \ \mbox{id} \ \|_{C^1} \ra 0\ .
\]
Let 
\begin{eqnarray}
\th &=& \inf_{u,v,w,q_j\in S_j,j\in {\mathbb Z}} \bigg \{ \min \{ 
\|u-v\|,\|v-w\|,\|w-u\|\} \ \bigg | \ u \in T_{q_j}\F^u(q_-)\ , \nonumber \\
& &  v \in T_{q_j}\F^s(q_+)\ ,  \ w \in T_{q_j}S_j\ ,\ \|u\|=\|v\|=\|w\|=1 
\bigg \}\ , \nonumber
\end{eqnarray}
where $T_{q_j}$ indicates the tangent space at $q_j$.
In this article, transversality always implies that such angle $\th$ 
is positive.
\item {\bf (A2)}. Let $\Om$ be a neighborhood of $S$. Then there exists 
a $J>0$ such that $S_j \subset \Om$, $\forall |j| \geq J$. Let
\[
d = \inf_{q \in S_j \cup S, |j| \geq J}\{ \ \mbox{distance}\ \{ q, \pa 
\Om \}\}\ ,
\]
then $d >0$. Let $\Om_j$ be a neighborhood of $S_j, \forall |j| < J$, 
\[
d_j = \inf_{q \in S_j}\{ \ \mbox{distance}\ \{ q, \pa \Om_j \}\}\ ,
\]
then $d_j >0$. The collection $B_\xi = \{ \Om, \Om_j \ |j| < J\}$ is 
called a tubular neighborhood of $\xi \cup S$. For any $0<n<\infty$, 
there exists such a tubular neighborhood $B_\xi$ of $\xi \cup S$, such 
that for any $q_1 \in B_\xi$ there is a $q \in \xi \cup S$, $q_1$ and 
$q$ belong to the same $\Om$ or $\Om_j$ $|j| < J$, 
\[
\| D^\ell F^{\pm n}(q_1) -  D^\ell F^{\pm n}(q)\| <1\ , \quad (\ell =1,2)\ .
\]
Moreover, 
\[
\max_{+,-}\sup_{q \in \xi \cup S} \| D^2F^{\pm n}(q)\| < \infty\ .
\]
Let
\[
\La_\ell = \max_{+,-}\sup_{q \in B_\xi} \| D^\ell F^{\pm n}(q)\| < \infty\ , 
\quad (\ell =1,2)\ .
\]
\end{itemize}
If $\xi \cup S$ is compact, then assumption {\bf (A2)} holds \cite{Li03}. 
Next we will introduce a pseudo-invariant tube. An invariant tube is a sequence of submanifolds $\{ B_j \}_{j \in {\mathbb Z}}$ such that
\[
F(B_j)=B_{j+1}\ , \quad F^{-1}(B_j) = B_{j-1}\ , \quad \forall j \in {\mathbb Z}\ . 
\]
\begin{definition}
Segment-0 is defined as the finite sequence of 2N+1 $S$'s
\[
\eta_0 = (S \cdots S)\ ,
\]
and Segment-1 is defined as the finite sequence
\[
\eta_1 = (S_{-N} S_{-N+1} \cdots S_0 \cdots S_{N-1} S_N)\ ,
\]
where $N$ is a large positive integer.
\end{definition}
\begin{definition} 
Let $\Sigma $ be a set that consists of
elements of the doubly infinite sequence form:
\[
a=(\cdots a_{-2}a_{-1}a_{0}, a_{1}a_{2}\cdots ),
\]
where $a_{k}\in \{0,1\}$, $k\in {\mathbb Z}$. We introduce a topology
in $\Sigma$ by taking as neighborhood basis of
\[
a^{*}=(\cdots a^{*}_{-2}a_{-1}^{*}
a^{*}_{0},a^{*}_{1}a^{*}_{2}\cdots ),
\]
the set
\[
\A_{j}=\{a\in \Sigma \ |\ \ a_{k}=a^{*}_{k} \ (|k|<j)\}
\]
for $j=1,2,\cdots $. This makes $\Sigma$ a topological space. The
Bernoulli shift automorphism $\chi$ is defined on $\Sigma$ by
\[
\chi :\Sigma \mapsto \Sigma ,\quad \forall a\in \Sigma ,\ 
\chi (a)=b, \ \mbox{where}\ b_{k}=a_{k+1}.
\]
\end{definition}
The Bernoulli shift automorphism $\chi$ 
exhibits sensitive  dependence on initial conditions,
which is a hallmark of chaos.

For any $\dl >0$, there exists $N>0$, such that
\[
\| \vphi^+_j - \ \mbox{id}\ \|_{C^1} < \dl \ , \quad 
\| \vphi^-_{-j} - \ \mbox{id}\ \|_{C^1} < \dl \ , \quad 
\forall j \geq N \ ,
\]
by assumption {\bf (A1)}.
\begin{definition}
To each $a_{k}\in \{ 0,1\}$, we associate the Segment-$a_{k}$,
$\eta_{a_{k}}$. Then each doubly infinite sequence
\[
a=(\cdots a_{-2}a_{-1}a_{0},a_{1}a_{2}\cdots )
\]
is associated with a $\dl$-pseudo-invariant tube
\[
\eta_{a}=(\cdots
\eta_{a_{-2}}\eta_{a_{-1}}\eta_{a_{0}},\eta_{a_{1}}\eta_{a_{2}}\cdots ).
\]
\label{PIT}
\end{definition}

\section{Fenichel Fiber Coordinates and a $\la$-Lemma}

\subsection{Fenichel Fiber Coordinates}

We will introduce Fenichel fiber coordinates in a neighborhood of $S$. 
For any $\th \in S$, let
\[
E^u(\th) =T_\th{\mathcal F}^u(\th), \quad 
E^c(\th) =T_\th S            , \quad 
E^s(\th) =T_\th {\mathcal F}^s(\th), 
\]
where $T_\th$ indicates the tangent space at $\th$. $E^u$ and $E^s$ 
provide a coordinate system for a neighborhood of $S$, that is, any point 
in this neighborhood has a unique coordinate 
\[
(\tv^s,\tth,\tv^u), \quad \tv^s \in E^s(\tth), \quad \tv^u \in E^u(\tth),
\quad \tth \in S.
\]
Fenichel fibers provide another coordinate system for the 
neighborhood of $S$. For any $\th \in S$, the Fenichel fibers ${\mathcal F}^s(\th)$
and ${\mathcal F}^u(\th)$ have the expressions
\[
\left \{ \begin{array}{l} \tv^s = v^s , \\ \tth= \th +\Th_s(v^s,\th), \\
\tv^u = V_s(v^s,\th), \end{array} \right.
\quad \quad \mbox{and}\quad \quad 
\left \{ \begin{array}{l} \tv^u = v^u , \\ \tth= \th +\Th_u(v^u,\th), \\
\tv^s = V_u(v^u,\th), \end{array} \right.
\]
where $v^s$ and $v^u$ are the parameters parametrizing ${\mathcal F}^s(\th)$
and ${\mathcal F}^u(\th)$,
\[
\Th_z(0,\th)= \frac {\pa} {\pa v^z}\Th(0,\th) = V_z(0,\th)= 
\frac {\pa} {\pa v^z}V_z(0,\th)= 0, \quad z=u,s;
\]
and $\Th_z(v^z,\th)$ and $V_z(v^z,\th)$ ($z=u,s$) are $C^3$ in $v^z$ 
and $C^2$ in $\th$. The coordinate transformation from ($v^s,\th,v^u$) 
to ($\tv^s, \tth,\tv^u$)
\[
\left \{ \begin{array}{l} \tv^s = v^s +V_u(v^u,\th), \\ 
\tth= \th +\Th_u(v^u,\th)+\Th_s(v^s,\th), \\
\tv^u =v^u+ V_s(v^s,\th), \end{array} \right.
\]
is a $C^2$ diffeomorphism. In terms of the Fenichel coordinate 
($v^s,\th,v^u$), the Fenichel fibers coincide with their tangent 
spaces. From now on, we always work with the Fenichel coordinate 
($v^s,\th,v^u$).

\subsection{$\la$-Lemma}

$\forall j \in {\mathbb Z}$ and $\forall q_j \in S_j$, $q_j$ is on 
a unique stable fiber $\F^s(q_+)$, $q_+ \in S$, and a unique unstable 
fiber $\F^u(q_-)$, $q_- \in S$. Let
\[
E^u(q_j) =T_{q_j}{\mathcal F}^u(q_-), \quad 
E^c(q_j) =T_{q_j}S_j            , \quad 
E^s(q_j) =T_{q_j}{\mathcal F}^s(q_+). 
\]
By assumption {\bf (A1)}, $E^u(q_j)$, $E^c(q_j)$, and $E^s(q_j)$ are 
$C^2$ in $q_j \in S_j$.
\begin{lemma}[$\la$-Lemma]
For any $\e >0$, there exists a $J >0$ such that
\begin{enumerate}
\item When $j \geq J$, $E^u(q_j)\oplus E^c(q_j)$ is $\e$-close to 
$E^u(q_+)\oplus E^c(q_+)$.
\item When $j \leq -J$, $E^s(q_j)\oplus E^c(q_j)$ is $\e$-close to 
$E^s(q_-)\oplus E^c(q_-)$.
\end{enumerate}
\end{lemma}

Proof: When $j$ ($>0$) is large enough, $q_j$ is in a neighborhood of 
$S$ where $\F^s(q_+) = E^s(q_+)$. Let $v_1 \in E^u(q_j)\oplus E^c(q_j)$,
$\| v_1 \| = 1$. One can represent $v_1$ in the frame 
($E^s(q_+), E^u(q_+)\oplus E^c(q_+)$),
\[
v_1 = (v_1^s, v_1^{uc})\ .
\]
Let $\la_1 = \| v_1^s\| /\| v_1^{uc} \|$. By assumption {\bf (A1)}, 
transversality implies that $\la_1$ has an absolute upper bound. The 
rest of the argument is the same with that in \cite{Li03}. Q.E.D.

\subsection{A Rectification}

Next we conduct a rectification in a neighborhood of $S$, which is 
necessary for graph transform argument later on. When $j$ ($>0$) is 
large enough, $q_j$ is in a neighborhood of $S$ where $\F^s(q_+) = 
E^s(q_+)$. Thus $E^s(q_j)= E^s(q_+)$. For any $\tv^u \in E^u(q_+)$, 
$\| \tv^u \| = 1$, $\tv^u$ has the representation in the frame 
($E^s(q_j), E^u(q_j)\oplus E^c(q_j)$),
\[
\tv^u = v^s + v^{uc}\ .
\]
All such $v^{uc}$'s span the projection $\tE^u(q_j)$ of $E^u(q_+)$ 
onto $E^u(q_j)\oplus E^c(q_j)$. $E^u(q_j)\oplus E^c(q_j)$ is $C^2$ in 
$q_j$. Shifting $E^u(q_+)$, $E^c(q_+)$, and $E^s(q_+)$ to $q_j$, they 
are also $C^2$ in $q_j$. Representing $E^u(q_j)\oplus E^c(q_j)$ in the 
frame ($E^u(q_+), E^c(q_+), E^s(q_+)$), $\tE^u(q_j)$ can be obtained 
from $E^u(q_j)\oplus E^c(q_j)$ by restricting $\th =0$, where $\th$ 
coordinatizes $E^c(q_+)$. Thus $\tE^u(q_j)$ is also $C^2$ in $q_j$. 
The rectification amounts to replacing $E^u(q_j)$ by $\tE^u(q_j)$. We 
will use the same notation $E^u(q_j)$. Similarly, when $j$ ($>0$) is 
large enough, one can rectify $E^s(q_{-j})$ inside $E^s(q_{-j})\oplus 
E^c(q_{-j})$.

\section{Shadowing Lemma and Chaos}

Let $\eta_a$ be a $\dl$-pseudo-invariant tube defined in Definition \ref{PIT}, 
\[
\eta_a = (\cdots \tS_{-1} \tS_0 \tS_1 \cdots )
\]
where $\tS_j = S_k$ or $S$; $j,k\in {\mathbb Z}$. Denote by 
$\widehat{E}$ the transversal bundle
\[
\widehat{E}=\{ (q, E^u(q), E^s(q))\ | \ q \in \eta_a \} ,
\]
which serves as a coordinate system around $\eta_a$, with the coordinate  
denoted by
\[
(q, x^u, x^s), \quad \mbox{where}\ q \in \eta_a, \quad x^u \in E^u(q), 
\quad x^s \in E^s(q).
\]
In this coordinate system, the map $F^n$ has the representation
\[
F^n(q, x^u, x^s)=(f(q, x^u, x^s), g^u(q, x^u, x^s), 
g^s(q, x^u, x^s)),
\]
where $n$ is a large positive integer. If $q \in \tS_j$, then $f(q, x^u, x^s)
\in \tS_{j+n}$.
\begin{lemma}
$\forall \mu >0$, fix a $n$ large enough, 
and fix a $\e$ small enough, if $\dl$ is sufficiently small, then
\begin{eqnarray}
& & (\La_1)^k \Pi_3^s <\frac{1}{2},\ \ (0\leq k\leq 2), \quad 
\Pi_\ell^s < \mu ,\ \ (\ell =1,2), \nonumber \\ 
& & (\La_1)^k \widehat{\Pi}_2^u  <\frac{1}{2},\ \ (0\leq k\leq 2),
 \quad \Pi_\ell^u <\mu , \ \  (\ell =1,3), \nonumber
\end{eqnarray}
where $\| x^u\| \leq \epsilon$, $\| x^s\|\leq 
\epsilon $, $D_1=D_q$, $D_2=D_{x^u}$,
$D_3=D_{x^s}$, and 
\begin{eqnarray}
& & \Pi ^s_\ell =\sup_{q ,x^u,x^s}\| D_\ell g^s(q ,x^u,x^s)\|, \ \ 
(\ell =1,2,3), \nonumber \\ 
& & \Pi^u_\ell =\sup_{q ,x^u ,x^s}\| D_\ell 
g^u(q ,x^u,x^s)\| , \ \ (\ell =1,2,3), \nonumber \\
& & \widehat{\Pi}^u_2=\sup_{q ,x^u,x^s}\| \{D_2g^u(q ,x^u,x^s)\}^{-1}\|. 
\nonumber
\end{eqnarray}
\label{est} 
\end{lemma}

Proof: The proof of this lemma follows from assumptions {\bf (A1)} and 
{\bf (A2)}, and the fact that along Segment-0 and Segment-1, the 
center-unstable and center-stable bundles $E^u(q)\oplus E^c(q)$ and 
$E^s(q)\oplus E^c(q)$ are invariant under the linearized flow \cite{Li03}.
Q.E.D.

Let $\Gamma_\epsilon$ be the space of sections of 
$\widehat{E}$:
\[
\Gamma_\epsilon = \{ \sigma \ | \ \sigma (q) =(q 
,x ^u(q), x ^s(q)), q \in \eta_a, \| x ^u(q ) \|
\leq \epsilon ,\| x^s(q) \| \leq \epsilon \}.
\]
We define the $C^0$ norm of $\sigma \in \Gamma_\epsilon$ as
\[
\| \sigma \|_{C^0}=\max \{ \sup_{q \in \eta_a}\| 
x^u(q )\| , \sup_{q
\in \eta_a}\| x^s(q )\| \}.
\]
Then we define a Lipschitz semi-norm on $\Gamma_\epsilon$:
\[
\mbox{Lip}\ \{\sigma \} = \max \left\{ 
\sup_{\|q _1-q _2\| \leq \Delta} \frac{\| 
x^u(q_1)-x^u(q
_2)\| }{\|
q_1-q_2\|}\right.\left. ,\sup_{\|q_1-q_2\| \leq 
\Delta}\frac{\| x^s(q _1)-x^s(q
_2)\|}{\|q_1-q_2\|}\right \}
\]
for some small fixed $\Delta >0$. Let $\Gamma_{\epsilon, \ga }$ be a subset of $\Ga_\e$, 
\[
\Gamma_{\epsilon, \ga }=\{ \sigma \in \Gamma_\epsilon 
\ |\ \ \ \mbox{Lip}\ \{ \sigma \} \leq \ga \}.
\]
For any $\sigma \in \Gamma_{\epsilon , \ga }$,
\[
\sigma (q)=(q 
,x^u(q),x^s(q)),\quad q \in \eta_a,
\]
we define the {\em{graph  transform}} $G$ as follows:
\begin{equation}
(G\sigma )(q)=(q, x^u_1(q), x^s_1(q)),
\label{gt} 
\end{equation}
where
\begin{eqnarray}
& & f(q^-,x^u(q^-),x^s(q^-))=q , \nonumber \\
& & g^s(q^-,x^u(q^-),x^s(q^-))=x^s_1(q), \nonumber \\ 
& & f(q ,x^u_1(q),x^s(q))=q^+, \nonumber \\ 
& & g^u(q,x^u_1(q),x^s(q))=x^u(q^+), \nonumber
\end{eqnarray}
for some $q^-$ and $q^+$.
\begin{theorem}
The graph transform $G$ is a contraction map on $\Gamma_{\epsilon ,\gamma}$.
The graph of the fixed point $\sigma^*$ of $G$ is an invariant tube under $F$.
\end{theorem}

Proof: The proof of this theorem is similar to that given in details in 
\cite{Li03}. Q.E.D.

Graph$\{\sigma^*\}$ is the invariant tube that $\e$-shadows the 
$\dl$-pseudo-invariant tube $\eta_a$. Let $S^*$ be the element of 
Graph$\{\sigma^*\}$, that shadows the mid-element of $\eta_a$, which 
is either $S_0$ or $S$. Such $S^*$'s for all $\eta_a$ forms a set $\Xi$ 
of submanifolds. It is obvious that the following theorem holds.
\begin{theorem}[Chaos Theorem]
The set $\Xi$ of submanifolds is invariant under the map $F^{2N+1}$. 
The action of $F^{2N+1}$ on $\Xi$ is topologically conjugate to the action of 
the Bernoulli shift automorphism $\chi$ on $\Sg$. That is, there exists a 
homeomorphism $\phi: \Sg \mapsto \Xi $ such that the following diagram 
commutes:
\[
\begin{array}{ccc}
\Sg &\maprightu{\phi} & \Xi \\
\mapdownl{\chi} & & \mapdownr{F^{2N+1}}\\
\Sg & \maprightd{\phi} & \Xi
\end{array} 
\]
\label{bshift}
\end{theorem}

\end{document}